\documentclass[pra,aps,twocolumn,nopacs,superscriptaddress,nofootinbib]{revtex4}
\usepackage{graphicx}
\usepackage{dcolumn}
\usepackage{bm}
\usepackage{amsmath}
\usepackage{epsfig}
\usepackage{color}

\def\ii{{\rm i}}  \def\ee{{\rm e}} \def\de{d}

\def\rb{{\bf r}}  \def\Rb{{\bf R}}    \def\vb{{\bf v}}
\def\xx{\hat{\bf x}}  \def\yy{\hat{\bf y}}

\def\kb{{\bf k}}    
    
\def\me{m_{\rm e}}  \def\kB{{k_{\rm B}}}
\def\Eb{{\bf E}}      
  \def\jb{{\bf j}}
  
\def\vF{v_{\rm F}}    \def\EF{{E_{\rm F}}}

\begin{document}
\title{Circular Dichroism in Rotating Particles}
\author{Deng~Pan}
\email[Corresponding author: ]{deng.pan@icfo.eu}
\affiliation{ICFO-Institut de Ciencies Fotoniques, The Barcelona Institute of Science and Technology, 08860 Castelldefels (Barcelona), Spain}
\affiliation{School of Physics and Technology, Wuhan University, Wuhan 430072, China}
\author{Hongxing~Xu}
\affiliation{School of Physics and Technology, Wuhan University, Wuhan 430072, China}
\author{F.~Javier~Garc\'{\i}a~de~Abajo}
\email[Corresponding author: ]{javier.garciadeabajo@nanophotonics.es}
\affiliation{ICFO-Institut de Ciencies Fotoniques, The Barcelona Institute of Science and Technology, 08860 Castelldefels (Barcelona), Spain}
\affiliation{ICREA-Instituci\'o Catalana de Recerca i Estudis Avan\c{c}ats, Passeig Llu\'{\i}s Companys 23, 08010 Barcelona, Spain}

\begin{abstract}
\textbf{Light interaction with rotating nanostructures gives rise to phenemona as varied as optical torques and quantum friction. Here we reveal that circular dichroism of rotating optically-isotropic particles has an unexpectedly strong dependence on their internal geometry. In particular, nanorings and nanocrosses exhibit a splitting of $2\Omega$ in the particle optical resonances, while compact particles display weak circular dichroism at low rotation frequency $\Omega$, but a strong circular dichroism at high $\Omega$. We base our findings on a quantum-mechanical description of the polarizability of rotating particles, which has not been rigorously addressed so far. Specifically, we use the random-phase approximation and populate the particle electronic states according to the principle that they are thermally equilibrated in the rotating frame. We further provide insight into the rotational superradience effect and the ensuing optical gain, originating in population inversion as regarded from the lab frame, in which the particle is out of equilibrium. Surprisingly, we find the optical frequency cutoff for superradiance to deviate from the rotation frequency $\Omega$. Our results unveil a rich, unexplored phenomenology of light interaction with rotating objects.}
\end{abstract}
\date{\today}
\maketitle

\section{Introduction}

Intriguing chirality-dependent phenomena emerge during the interaction between circularly-polarized light and rotating objects. For example, the rotational Doppler effect \cite{G1981,CDR98,CRD98,CLH06,LSB13,KSG13} causes the emission of right- and left-circularly polarized (RCP and LCP) light from a particle rotating at frequency $\Omega$ to experience opposite frequency shifts $\pm\Omega$ \cite{G1981,CDR98}. As a consequence of this, inelastic scattering from an anisotropic rotating particle presents a chiral asymmetry consisting of $\pm2\Omega$ frequency shifts for RCP and LCP light \cite{G1981,LSB13}, also observed through rotational Raman scattering \cite{G1981,KSG13}. Chiral effects are found as well in elastic scattering by isotropic rotating media, such as the rotational photon drag (i.e., the rotation of polarization upon light transmission \cite{PWG06}, which can be enhanced in a slow-light medium \cite{FGB11}). In the extreme situation when the rotation frequency exceeds the optical frequency of the incident circularly-polarized light, losses in the medium turn into gains, which essentially produce light amplification at the expense of mechanical energy \cite{GF1947,Z1972,BS98,paper172}. This rotational superradiance effect plays a fundamental role in astrophysics, finding its counterpart in Hawking's radiation \cite{H1971} when dealing with rotating Kerr black holes \cite{B1973}. Additionally, rotation-induced chirality produces a friction torque upon interaction with the thermal or zero-point fluctuations of the vacuum electromagnetic field \cite{paper157,paper156,paper199,MJK12,BL15,paper289,paperPan}.

These phenomena can be described through a frequency-dependent polarizability $\alpha(\omega)$ for objects that are small compared with the wavelength. One might naively use values of $\alpha(\omega)$ obtained from the particle at rest and apply them to the rotating particle by a simple coordinate transformation, ignoring changes in the internal states that are produced by rotation. However, such widely-used prescription \cite{paper157,paper166,paper199,MJK12,MGK13,MGK13_2,MJK14,LS16,paper289} does not produce the correct frequency shifts, as we demonstrate in this paper. A more rigorous approach to model $\alpha(\omega)$ quantum mechanically then becomes necessary. However, a quantum description of the particle optical response including the effect of mechanical rotation is far from trivial, as even the ground state of rotating objects require elaborate computations \cite{W12,LGY12,B13,WO15,SSH18,SGH18}.

\begin{figure}[t]
\begin{centering}
\includegraphics[width=0.5\textwidth]{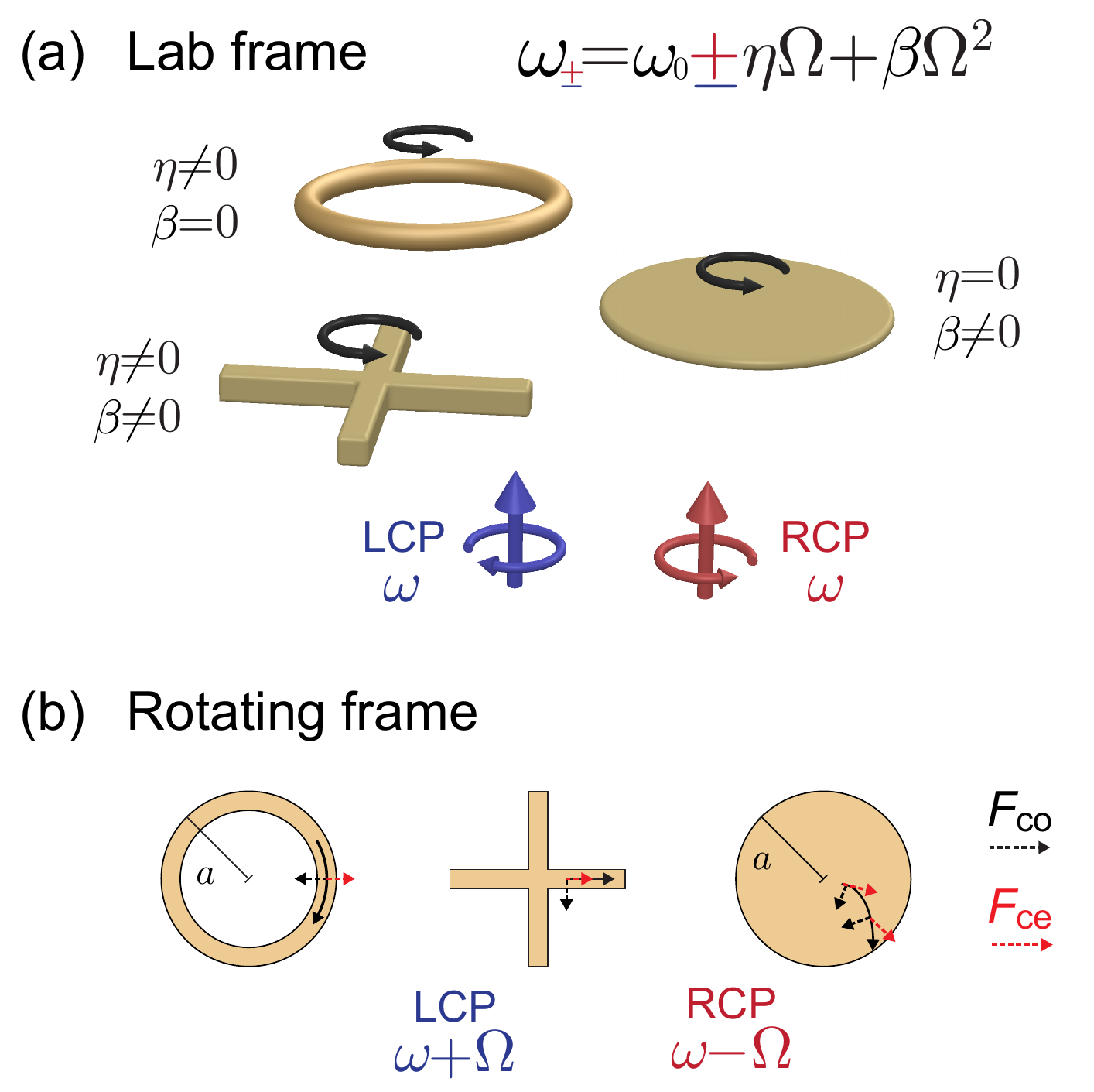}
\par\end{centering}
\caption{We consider optically isotropic particles characterized by a resonance frequency $\omega_0$ when they are at rest. (a) When the particles rotate with angular velocity $\Omega$, the optical resonance frequency becomes $\omega_\pm=\omega_0\pm\eta\Omega+\beta\Omega^2$ for RCP (+) and LCP ($-$) illumination in the lab frame (both equal to $\omega_0$ at $\Omega=0$), where the shape-dependent $\eta$ and $\beta$ terms are corrections due to Coriolis and centrifugal forces, respectively. (b) In the frame rotating with the particles, internal electrons experience Coriolis ($F_{\rm co}$) and centrifugal ($F_{\rm ce}$) forces, while the apparent frequency of circularly polarized incident light shifts to $\omega\mp\Omega$. The effect of $F_{\rm co}$ and $F_{\rm ce}$ depends on particle morphology.}
\label{Fig1}
\end{figure}

In this work, we study the optical response of rotating particles using the random-phase approximation (RPA) to calculate their response including quantum mechanical effects. In particular, we obtain optical polarizabilities for disks and rings rotating around their symmetry axis. Despite the fact that both types or particles share an optical isotropy in the plane perpendicular to the rotation axis, their optical responses toward circularly-polarized light shows strikingly different behaviors characterized by a resonance splitting by $\pm\Omega$ in rings and no splitting in disks at small finite rotation frequency $\Omega$. In classical terms, we attribute this discrepancy to the blocking of the Coriolis force in rings due to the constrained 1D motion of internal electrons along their circumference, in contrast to 2D electron motion in the disks. From a quantum-mechanical perspective, the effect is equivalently explained from the change in electronic state energies in the frame rotating with the particle, where thermal equilibrium is internally established. Further intriguing conclusions are established based on our model on the optical frequency cutoff for superradiance, which we find to significantly deviate from $\Omega$. These results open unexpected perspectives on the optical response of rotating objects.

\section{Dichroism of rotating particles}

We consider particles that are optically isotropic for polarization perpendicular to $z$ (Fig.\ \ref{Fig1}). Two degenerate dipolar modes can then be individually excited with polarization $p_\pm(\omega)=p_x(\omega)\pm \ii p_y(\omega)$ in the perpendicular plane in response to the electric field $E_\pm(\omega)=E_x(\omega)\pm \ii E_y(\omega)$ of RCP (+) and LCP ($-$) light incident along $z$ [bottom arrows in Fig. 1(a)]. Assuming linear response, we have $p_\pm(\omega)=\alpha_\pm(\omega) E_\pm(\omega)$, where $\alpha_\pm(\omega)$ are the corresponding circular polarizability components of the particle. In the absence of rotation ($\Omega=0$), the particle is chirally symmetric, $\alpha_+^0(\omega)=\alpha_-^0(\omega)=\alpha^0(\omega)$. In what follows, we consider a response dominated by electronic excitations and characterized by a spectrally isolated resonance $\omega_0$ showing up as a degenerate peak in ${\rm Im}\{\alpha_\pm^0(\omega)\}$. Nonetheless, our results can be trivially extended to other types of modes, for example of phononic nature.

When the particle rotates at a frequency $\Omega$ around the $z$ axis, the resonance frequencies of the two dipolar modes $p_\pm(\omega)$ are shifted to $\omega_\pm=\omega_0\pm\eta \Omega+\beta\Omega^2$, where $\eta$ and $\beta$ depend on particle geometry (see below), and the sign in front of $\eta$ denotes an optical CD. To intuitively understand these shifts, we view the system within the frame rotating with the particle [Fig.\ \ref{Fig1}(b)], in which particle electrons experience Coriolis and centrifugal forces, $F_{\rm co}$ and $F_{\rm ce}$, respectively. The Coriolis force is a classical manifestation of Berry's phase, closely related to the magnetization of materials by rotation \cite{B1915,B1935_2,H1962} and topological nontrivial phenomena \cite{S95,W15}. The direction of $F_{\rm co}$ depends on the sign of $\Omega$, giving rise to a correction $\pm\eta\Omega$, where $\eta$ determines the magnitude of the resulting CD of the particle. In contrast, the centrifugal force is always oriented along the outward radial direction, and therefore, its contribution is independent of the sign of rotation, leading to the quadratic $\beta\Omega^2$ correction. Now, in a thin nanoring, the internal electrons are confined to move along the ring circumference, so their motion is not affected by the forces $F_{\rm co}$ and $F_{\rm ce}$ [Fig.\ \ref{Fig1}(b)]; in consequence, the nanoring polarizability observed in the rotating frame must coincide with the polarizability of the motionless ring $\alpha^0(\omega)$, characterized by a $\omega_0$ resonance. However, the coordinate transformation associated with rotation makes the frequency $\omega$ of RCP and LCP light in the lab to appear as $\omega\mp\Omega$ in the rotating frame [bottom arrows in Figs.\ \ref{Fig1}(a) and \ref{Fig1}(b)], thus Doppler-shifting the resonance captured by $p_\pm(\omega)$ to $\omega_\pm=\omega_0\pm \Omega$ in the lab frame. Similar considerations also lead to a $2\Omega$ resonance splitting in the lab frame for a rotating thin nanocross, where geometrical confinement disables $F_{\rm co}$, although $F_{\rm ce}$ is active and produces identical shifts in both frequencies $\omega_\pm$. In contrast, in an extended particle such as a disk [Fig.\ \ref{Fig1}(b)], both $F_{\rm co}$ and $F_{\rm ce}$ affect the motion of electrons, leading to a more complex dependence of $\alpha(\omega)$ on rotation, as revealed by our detailed calculations presented below; in particular, in a free-electron description of a disk, the Coriolis force completely compensates the rotational Doppler effect, rendering $\eta=0$ (i.e., no CD). 
In a disk rotating at high frequency the centrifugal force $F_{\rm ce}$ tends to push the unperturbed electron density toward the particle edge, thus leading to a ring-like configuration and enhancing the CD.

\section{Quantum description of rotating particles}

We model the circular polarizabilities $\alpha_\pm(\omega)$ of nanorings and nanodisks in the random-phase approximation (RPA) \cite{HL1970} (see Appendix). Due to axial symmetry, each internal one-electron state of the particle $\left|j\right\rangle$ (wave function $\psi_j$) has a well-defined azimuthal quantum number $m_j$. We neglect retardation in the internal description of the particle, and accordingly represent the external circularly polarized light through a scalar potential $\phi^{\rm ext}_\pm(\rb)=-E^{\rm ext}R\ee^{\pm\ii\varphi}/\sqrt{2}$, where we use cylindrical coordinates $\rb=(R,\varphi,z)$ and implicitly assume a time dependence $\ee^{-\ii\omega t}$. The charge distribution induced in the particle $\rho_\pm(\rb)=\int {\rm d}\rb'\chi(\rb,\rb')\phi_\pm(\rb')$ is related to the potential through the susceptibiliy $\chi(\rb,\rb')$, while the sought-after polarizabilities reduce to $\alpha_\pm(\omega)=\int{\rm d}\rb R\ee^{\mp\ii\varphi}\rho(\rb)/E^{\rm ext}$. Using matrix notation, we can write $\chi=\chi^0\cdot(1-v\cdot\chi^0)^{-1}$ in terms of the Coulomb interaction $v(\rb,\rb')=1/|\rb-\rb'|$ and the non-interacting susceptibility $\chi^0(\rb,\rb')$. In the RPA, the latter admits the expression \cite{HL1970}
\begin{align}
\chi^0(\rb,\rb')\!=\!\frac{2 e^2}{\hbar}\sum_{jj'} (f_{j'}-f_j)\frac{\psi_j(\rb)\psi^*_j(\rb')\psi^*_{j'}(\rb)\psi_{j'}(\rb')}{\omega-(E_j-E_{j'})/\hbar+\ii \gamma},
\label{chi0}
\end{align}
where $E_j$ and $f_j$ denote the energy and population of state $\left|j\right\rangle$, $\gamma$ is a phenomenological damping rate, and the factor of 2 accounts for spin degeneracy.

The particle rotation frequency $\Omega$ enters Eq.\ (\ref{chi0}) through the populations $f_j$. Incidentally, a recent proposal for a quantum time crystal based on the emergence of an observable that is rotating in the lab frame \cite{W12,LGY12,B13,WO15} has been proven to be realizable only if the system is out of thermal equilibrium. But here, we assume thermal equilibrium, which must be fulfilled in the rotating frame. In order to determine $f_j$, we thus need to transform the Schr\"odinger equation from the lab frame [coordinates $(R,\varphi,z,t)$] to the rotating frame [coordinates $(R',\varphi',z',t')=(R,\varphi-\Omega t,z,t)$], where it becomes $(\mathcal{H}_0-\Omega\mathcal{L}_z)\psi=\ii\hbar\partial_{t}\psi$. Here, $\mathcal{H}_0$ is the Hamiltonian of the motionless particle, while the term $-\Omega\mathcal{L}_z$ accounts for a quantum description of $F_{\rm co}$ and $F_{\rm ce}$. For particles with axial symmetry, the energies of internal quantum states $\left|j\right\rangle$ are shifted from $E_j$ in the motionless particle to $\tilde{E}_j=E_j-m_j\hbar\Omega$ in the rotating frame, and their populations described by the Fermi-Dirac distribution become $f_j=[\ee^{(\tilde{E}_j-\tilde{E}_{\rm F})/\kB T}+1]^{-1}$, where $\tilde{E}_{\rm F}$ is the Fermi energy in the rotating frame.

\begin{figure}[t]
\begin{centering}
\includegraphics[width=0.5\textwidth]{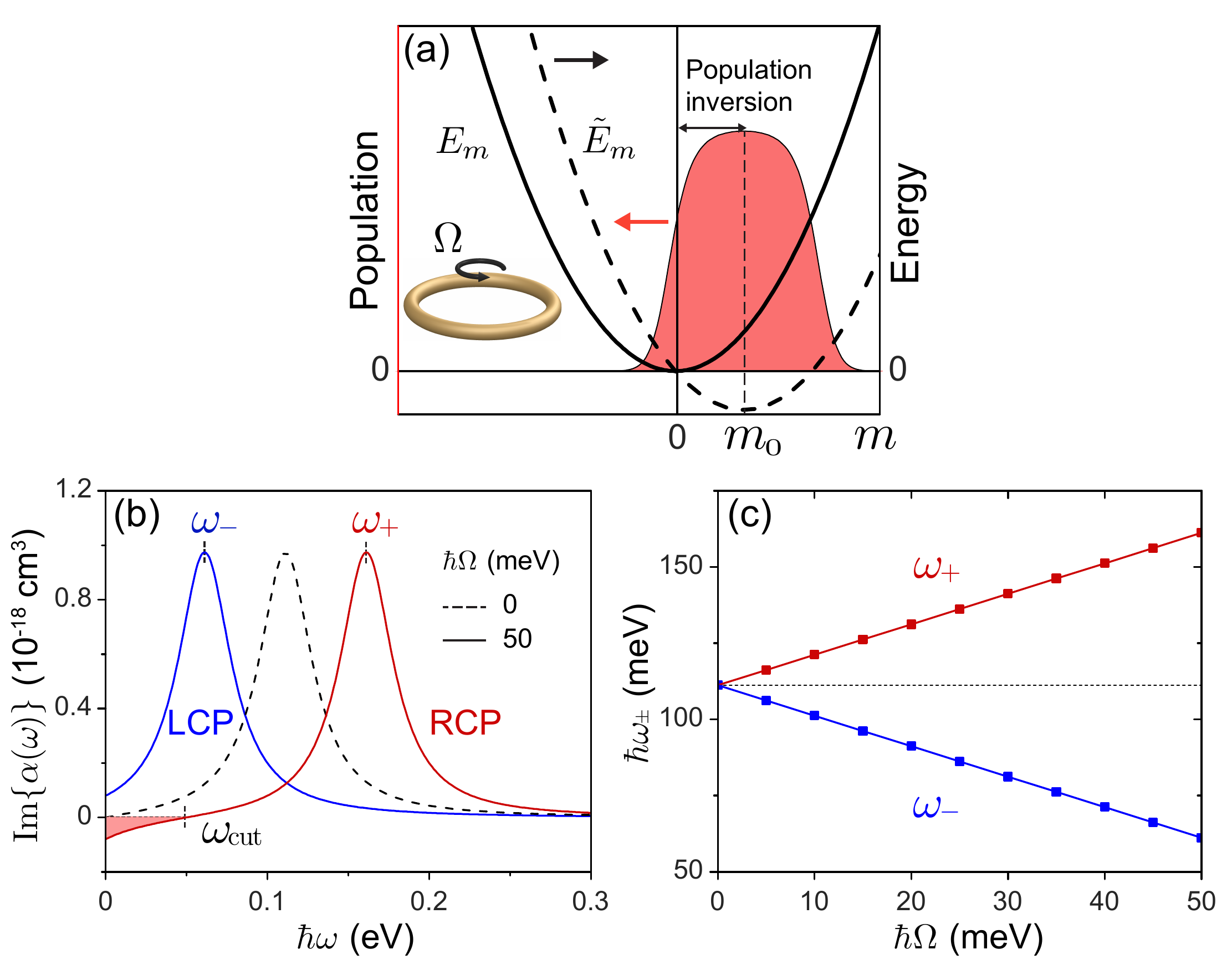}
\par\end{centering}
\caption{(a) Energy spectra (black curves) and thermal population (red shaded curve) of the electron states $\left|m\right\rangle$ in a quasi-one-dimensional nanoring. Solid and dash black curves correspond to electron states in the ring observed in the lab frame (energies $E_m$) and in a frame rotating at a frequency $\Omega$ (energies $\tilde{E}_m$), respectively. If the ring rotates at frequency $\Omega$, its internal states are in thermal equilibrium in the rotating frame, so the population $f_m$ of states $\left|m\right\rangle$ is determined by $\tilde{E}_m$. (b) Quantum-mechanical results for the LCP and RCP polarizability tensor components of a rotating nanoring (radius $a=8$\,nm, 80 electrons, damping $\hbar\gamma=20$\,meV) for $\hbar\Omega=50$\,meV. Results for the motionless particle (dashed curve) are displayed for comparison. The particle temperature is taken to be $300$\,K. (c) Position of the resonance frequencies for LCP and RCP as a function of rotation frequency.}
\label{Fig2}
\end{figure}

\section{Polarizability of rotating nanorings}

The population $f_j$ in a rotating nanoring is illustrated in Fig.\ \ref{Fig2}(a), and the resulting polarizabilities are plotted in Fig.\ \ref{Fig2}(b). We label electron states in a quasi-one-dimensional nanoring by the azimuthal quantum numbers $m$ (i.e., we neglect radial degrees of freedom) and write the associated energies in the lab frame as $E_m=\hbar^2m^2/2\me a^2$ [solid curve in Fig.\ \ref{Fig2}(b)]. If the nanoring rotates at frequency $\Omega$, the populations $f_m$ [Fig.\ \ref{Fig2}(a), red shaded area] are determined by the energies $\tilde {E}_m=E_m-m\hbar\Omega$ observed in the rotating frame [Fig.\ \ref{Fig2}(a), dashed curve], so the maximum population appears at the {\it ground state} $\left|m_0\right\rangle$ in that frame. The particle rotation is evident by the nonzero total angular momentum $2\sum_m f_m\,m\hbar$ in this distribution. It is interesting to note that a population inversion exists in the lab frame for electron states in the $0<m<m_0$ region. Stimulated emission from these states occurs under external illumination with suitable polarization, therefore leading to amplification of the light intensity.

Figure\ \ref{Fig2}(b) shows the polarizabilites $\alpha_\pm(\omega)$ of a rotating nanoring (radius $a=8$\,nm, $80$ electrons confined in its circumference), calculated in the RPA model discussed above. Their spectral profiles are just translations of the motionless polarizability $\alpha^0(\omega)$ [Fig.\ \ref{Fig2}(b), black-dash curve], $\alpha_\pm(\omega)=\alpha^0(\omega\mp\Omega)$. For $\Omega>0$ and RCP light (or equivalently for $\Omega<0$ and LCP light), we have ${\rm Im}\{\alpha_+(\omega)\}<0$ [Fig.\ \ref{Fig2}(b), red area] at frequencies $\omega$ below a cutoff $\omega_{\rm cut}=\Omega$. As the extinction cross-section is proportional to the imaginary part of the polarizability, a negative value of the latter indicates that the particle produces optical gain for the incident RCP light of frequency $\omega<\Omega$; this gain originates in the population inversion observed in Fig.\ \ref{Fig2}(a). Incidentally, the magnitude of $|{\rm Im}\{\alpha_+(\omega)\}|$ at low frequency scales with the damping rate $\gamma$ [see Eq.\ (\ref{chi0})] because this parameter determines how fast out-of-equilibrium electrons [Fig.\ \ref{Fig2}(a)] can undergo transitions accompanied by the emission of radiation. The rotating nanoring also shows an optical CD, the magnitude of which scales linearly with $\Omega$ in accordance with the intuitive picture in Fig.\ \ref{Fig1}, as illustrated in Fig.\ \ref{Fig2}(c), where we plot the resonance frequencies $\omega_\pm=\omega_0\pm\Omega$ for RCP and LCP light, so in fact a rotation frequency much smaller than the value of $50$\,meV used in Fig.\ \ref{Fig2}(b) should be enough to manifest CD. We also note that a clear CD splitting in the spectral profiles of magnitude $\Omega>\gamma$ should be feasible because $\Omega$ characterizes the collective motion of electrons, which is independent on the internal collision rate $\gamma$.
\begin{figure}
\begin{centering}
\includegraphics[width=0.5\textwidth]{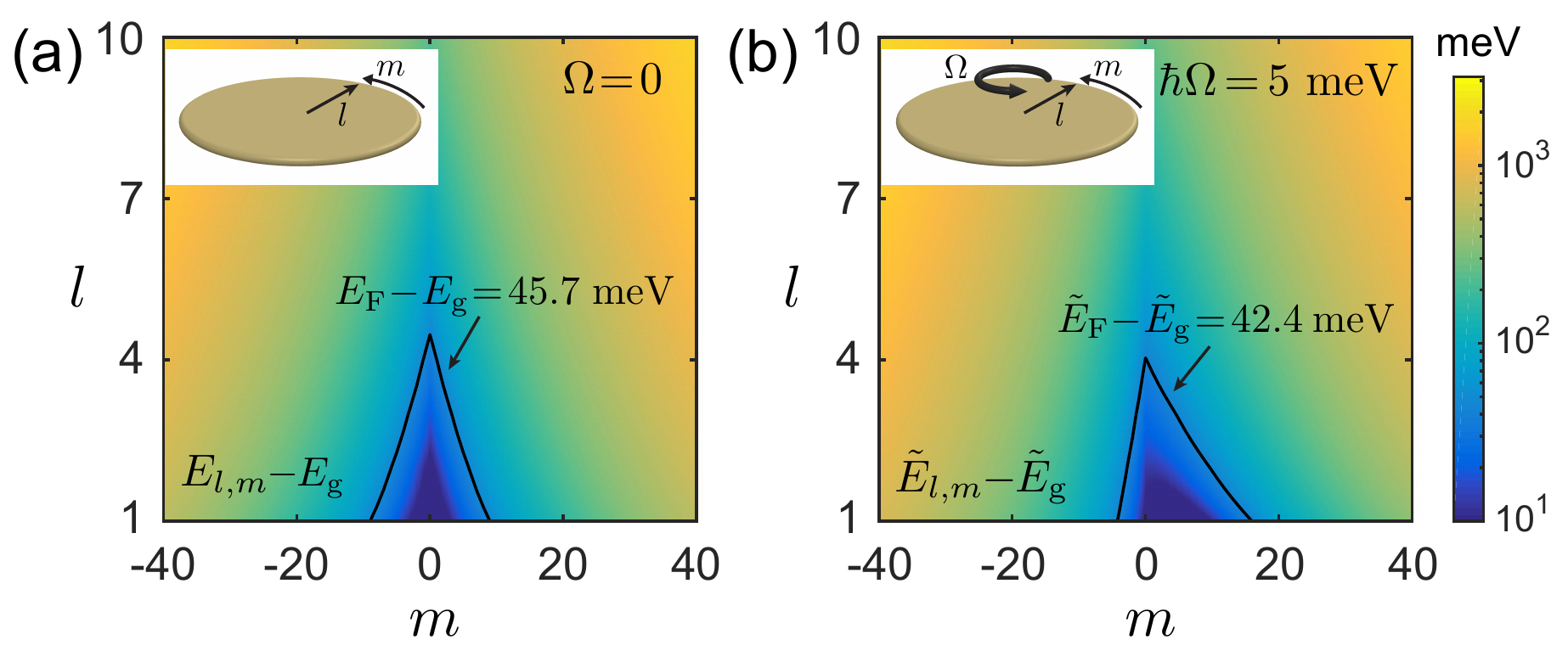}
\par\end{centering}
\caption{(a) Energy spectrum of a nanodisk (radius $a=12$\,nm) observed in the lab frame, where $m$ and $l$ denote azimuthal and radial quantum numbers. The black curve shows the Fermi level when the disk contains 80 electrons. (b) Energy spectrum in a rotating frame for the disk in (a) with $\hbar\Omega=5\,$meV. The ground state shifts from $|l,m\rangle=|1,0\rangle$ in the lab frame to $|1,4\rangle$ in the rotating frame. Energies are referred to the respective ground state energies $E_{\rm g}$ and $\tilde{E}_{\rm g}$.}
\label{Fig3}
\end{figure}

\begin{figure*}
\begin{centering}
\includegraphics[width=1\textwidth]{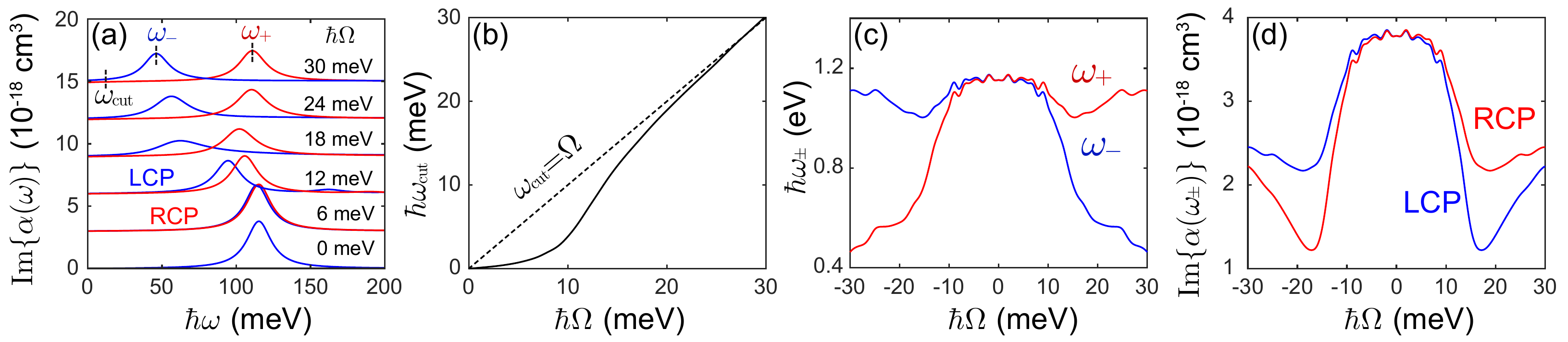}
\par\end{centering}
\caption{(a) Polarizability of a disk (radius $a=12$\,nm, 80 electrons, damping $\hbar\gamma=10$\,meV) rotating at different frequencies $\Omega$ (see labels) for LCP (blue) and RCP (red) light. (b) Rotation-frequency dependence of the cutoff for superradiance. (c,d) $\Omega$-dependence of the resonance frequency $\omega_\pm$ (c) and maximum polarizability (d) extracted from (a).}
\label{Fig4}
\end{figure*}

\section{Polarizability of rotating nanodisks}

For the nanodisk shown in the inset of Fig.\  \ref{Fig3}(a), in addition to the azimuthal quantum number $m$, the electron states are also labeled by a radial quantum number $l$, so that their energies in the motionless particle are $E_{lm}=\hbar^2\xi_{lm}^2/2a^2m_{\rm e}$, where $\xi_{lm}$ is the $l^{\rm th}$ zero of the Bessel function $J_m$. We assume a single out-of-plane electron state for simplicity. The energies $E_{lm}$ are represented in Fig.\ \ref{Fig3}(a) for a disk of radius $a=12$\,nm; the ground state is then $|l,m\rangle=|1,0\rangle$ ($E_{\rm g}=1.5$\,meV) and the Fermi energy (black curve) is $\EF=47.2$\,meV when the disk is filled with 80 electrons. The electron energies in a nanodisk observed in a frame rotating with frequency $\Omega$ become $\tilde{E}_{lm}=E_{lm}-m\hbar\Omega$. These energies are represented in Fig.\ \ref{Fig3}(b) for the same disk as in (a) and $\hbar\Omega=5$\,meV; the ground state energy ($\tilde{E}_{\rm g}=-4.8$\,meV) has now moved to $\left|1,4\right\rangle$ and the Fermi energy becomes $\EF=37.6$\,meV. If this disk is also rotating at frequency $\Omega$, the internal state populations $f_{lm}$ follow a Fermi-Dirac distribution determined by the rotating frame energies $\tilde{E}_{lm}$ as shown in Fig.\ \ref{Fig3}(b).

Figure\ \ref{Fig4}(a) shows the resulting polarizability ${\rm Im}\{\alpha(\omega)\}$ as computed from Eq.\ (\ref{chi0}) for different rotation velocities. Surprisingly, the cutoff frequency $\omega_{\rm cut}$ for superradience [Fig. \ref{Fig4}(b)] presents substantial deviations from $\Omega$, including its slope as a function of $\Omega$ at small rotation frequencies. Additionally, the resulting CD differs considerably in the rotating nanodisk compared with the rotating nanoring, as found upon inspection of the resonance frequencies $\omega_\pm$ [Fig.\ \ref{Fig4}(c)] and the magnitude at the peaks of ${\rm Im}\{\alpha(\omega)\}$ [Fig.\ \ref{Fig4}(d)]. At relatively low rotation frequencies ($|\hbar\Omega|<10$\,meV), the resonances $\omega_\pm$ deviate only slightly from $\omega_0$, so the resulting CD has a small magnitude. We explain this effect by considering the Coriolis force $F_{\rm co}$ in the rotation frame, which produces an effective magnetization on the particle, resulting in energy shifts $\mp\Omega$ that compensate the rotational Doppler shifts $\pm\Omega$. We further attribute small deviations from perfect compensation to the asymmetry of the Fermi energy in the rotating particle [i.e., the effect of a change in the populations $f_{lm}$ observed in Fig.\ \ref{Fig3}(b)].

The conservative centrifugal force defines an effective potential for the electrons, which tend to push the electron density toward the disk edge. Compared to the amount of electron energy gained from this effective potential, the Coulomb energy described by the self-consistent Hartree term can be disregarded due to the low electron density of the particle here studied, therefore resulting in an accumulation of electron density near the disk edge due to the centrifugal force captured in the $-\Omega\mathcal{L}_z$ term of the Hamiltonian in the rotating frame; at low rotation frequencies $|\hbar\Omega|<10$\,meV, this effect ($\beta$ term in Fig.\ \ref{Fig1}) produces an identical frequency shift in both resonance frequencies $\omega_\pm$. Additionally, it leads to a ring-like configuration (electrons confined near the edges), which becomes significant at high $\Omega$ (notice the $\Omega^2$ scaling of $F_{\rm ce}$), and its interplay with the Coriolis force results in a CD even stronger than in the ring ($|\omega_+-\omega_-|>2\Omega$). In addition to the quantum-mechanical approach here followed, all features of the polarizabilites of the rotating particles are corroborated by describing them within an extension of the Drude model that incorporates both Coriolis and centrifugal forces (see Appendix).

\section{Concluding remarks}

We have investigated the polarizability of optically isotropic particles through a quantum-mechanical approach that reveals a strong dependence of their response on the geometry of the particle. More precisely, the optical CD associated with the Doppler shift in rotating particles is either present or strongly canceled by Coriolis forces depending on whether material electrons are constrained to quasi-one-dimensional motion or follow a nearly-free evolution inside the particles. Nanorings and nanodisks illustrate extreme examples of these behaviors. Our formalism traces this finding back to the out-of-equilibrium distribution of the particle electronic state as observed in the lab frame, which affects the way in which they exchange energy and momentum with the surrounding vacuum. These processes are accompanied by a friction torque acting  on the particle, which has been the subject of recent investigations \cite{paper157,paper156,paper166,paper199,MJK12,BL15,MGK13,MGK13_2,MJK14,paper289,paperPan}, and on which our results shed new light by providing a more rigorous derivation of the particle polarizability that incorporates geometry-dependent Coriolis effects.

An experimental observation of the difference in CD between shell-(or ring-)like particles and solid particles should be attainable using currently available techniques. Considering that small particles have already been observed to rotate at gigahertz frequencies $\Omega$ in high vacuum \cite{RDH18,AXB18}, the exotic shape-dependent evolution of their CD under rotation leads to measurable splittings or their resonances. Narrow excitation lines such as those associated with dopant two-level emitters could be used for this purpose. Alternatively, doped graphene structures should facilitate the task because of their high electrical tunability resulting from the conical electron dispersion, and the long lifetime of the plasmons in this material \cite{NMS18} (see Appendix). Additionally, a graphene disk of $\sim 100$\,nm radius should mechanically resist high rotation frequencies \cite{LWK08}. In a different approach, a rotating particle could be mimicked by the electron currents generated by applying an electric voltage or also by those associated with the photon drag \cite{PWG06,FGB11} driven by a linearly polarized external light pump whose direction of polarization is rotating at a high frequency $\Omega$. As the cause of the predicted effect lies in the out-of-equilibrium particle distribution in the lab frame, an effective rotation could be also achieved through an electric current circulating around a graphene ring. Rotating quantum gases could undergo a similar CD that deserves further investigation.

\appendix

\section{Transformation of the particle Hamiltonian to the rotating frame}
We consider an axially symmetric particle rotating around its symmetry axis $z$ with angular frequency $\Omega$. Rotational symmetry allows us to choose a complete set of internal particle states $\left|j\right\rangle$ with well-defined azimuthal angular quantum numbers $m_j$. In what follows, we use cylindrical coordinates $\rb=(R,\varphi,z)$. The noninteracting linear susceptibility of the system at frequency $\omega$ in the random-phase approximation (RPA) reads \cite{PN1966}
\begin{align}
\chi^0(\rb,\rb')&=\frac{2e^2}{\hbar}\sum_{jj'}(f_{j'}-f_{j})\frac{\psi_j(\rb) \psi_j^*(\rb') \psi_{j'}^*(\rb) \psi_{j'}(\rb')}{\omega-(\varepsilon_j-\varepsilon_{j'})+\ii\gamma}, \label{Achi0}
\end{align}
where $\psi_j(\rb)$ are one-electron wave functions of energies $\hbar\varepsilon_j$ and occupations $f_j$, and we incorporate a phenomenological inelastic loss rate $\gamma$. Neglecting corrections beyond $\chi^0$, considering circularly polarized illumination of frequency $\omega$ with unit electric field $(\xx\pm \ii\yy)/\sqrt{2}$, and assuming the particle to be small enough to neglect retardation, the light can be described through an electric potential $\phi^{\rm ext}_\pm(\rb)=-R\ee^{\pm\ii \varphi}/\sqrt{2}$, which induces a charge density distribution in the particle given by
\begin{align}
\rho_\pm^{\rm ind}(\rb)&=\int d^3\rb'\chi^0(\rb,\rb')\,\phi^{\rm ext}_\pm(\rb') \nonumber \\
&=\frac{2e}{\hbar}\sum_{jj'}(f_{j'}-f_j)\frac{\psi_j(\rb) \psi_{j'}^*(\rb)}{\omega-(\varepsilon_j-\varepsilon_{j'})+\ii\gamma}p_{jj'}\delta_{m_j,m_j'\pm 1}, \nonumber
\end{align}
where we have made use of the transition dipole moments $p_{jj'}\delta_{m_j,m_j'\pm 1}=-e\langle \psi_j|R\ee^{\pm\ii \varphi}|\psi_{j'}\rangle/\sqrt{2}$ between states $|j\rangle$ and $|j'\rangle$, and rotational symmetry leads to $|\langle j|x|j'\rangle|=|\langle j|y|j'\rangle|=|p_{jj'}|/\sqrt{2}e$. Incidentally, an overall time-dependent factor $\ee^{-\ii\omega t}$ must be understood in the potential and charge density. The induced charge allows us to directly obtain the polarizability
\begin{align}
\alpha_\pm(\omega)&=\int d^3\rb \frac{R \ee^{\mp\ii \varphi}}{\sqrt{2}} \rho^{\rm ind}_\pm(\rb) \nonumber\\
&=\frac{2}{\hbar}\sum_{jj'}\frac{f_j-f_{j'}}{\omega-(\varepsilon_j-\varepsilon_{j'})+\ii\gamma}|p_{jj'}|^2\delta_{m_j,m_{j'}\pm 1} \label{alpha},
\end{align}
where the delta function reflects the fact that only transitions $m_j\rightarrow m_j+1$ or $m_j\rightarrow m_j-1$ take place for left (LCP) or right (RCP) circular polarization, respectively.

\section{Interacting RPA polarizability of rotating particles}
In the self-consistent RPA approach, the polarizability $\chi^0$ is taken to repond to the total potential $\phi_\pm(\rb)$ (i.e., the sum of external and induced potentials) instead of only the external potential $\phi_\pm^{\rm ext}(\rb)$. One can easily verify that the total potential and induced charge density admit the expressions $\phi_\pm=(1-v\cdot\chi^0)^{-1}\cdot\phi^{\rm ext}_\pm$ and $\rho_\pm^{\rm ind}=\chi^0\cdot\phi_\pm$, respectively, where we use matrix notation with space coordinates $\rb$ acting as matrix indices [i.e., a dot denotes matrix multiplication, or equivalently space integration, while $\rho_\pm^{\rm ind}$, $\phi_\pm^{\rm ext}$, and $\phi_\pm$ are vectors, $\chi^0$ and $v$ are matrices, and the components of the bare Coulomb interaction are $v(\rb,\rb')=1/|\rb-\rb'|$]. We can also define the interacting susceptibility as
\begin{align}
\chi=\chi^0\cdot(1-v\cdot\chi^0)^{-1},
\label{chi}
\end{align}
in terms of which the induced charge can be expressed as $\rho_\pm^{\rm ind}=\chi\cdot\phi^{\rm ext}_\pm$. Finally, the polarizability can be calculated from the induced charge distribution using Eq.\ (\ref{alpha}).

In what follows, we present details of the calculation of the RPA response for thin rings and disks. We implement the RPA numerically by using the electron wave functions of a ring- or or disk-like box potential, respectively, taking a finite number of occupied states. We further expand the Coulomb interaction in this basis sets, thus avoiding the integrable singularity at $\rb=\rb'$.
\subsection{Ring}
We consider a quasi-one-dimensional ring (radius $a$) of small cross section (width $b\ll a$) defined by an infinite potential well of ring-like shape, as illustrated in Fig.\ 1 of the main text. We assume $b$ to be small enough so that only the lowest-energy state [normalized transversal wave function $\psi_\perp(R,z)$] contributes to the response, and therefore, the one-electron wave functions can be written as $\psi_m(\rb)=\psi_\perp(R,z)\ee^{\ii m \varphi}/\sqrt{2\pi}$ and their energies $\hbar\varepsilon_m=m^2\hbar^2/2\me a^2$ referred to that of that transversal state without loss of generality. As a consequence of this simplifying assumption, we find from Eq.\ (\ref{Achi0}) that $\psi_\perp$ only enters $\chi_0$ through an overall multiplicative factor as $\chi^0(\rb,\rb')\propto|\psi_\perp(R,z)\psi_\perp(R',z')|^2$. We are ultimately interested in the interacting susceptibility (\ref{chi}), which admits the Taylor expansion $\chi=\chi_0+\chi_0\cdot v\cdot\chi_0+\chi_0\cdot v\cdot\chi_0\cdot v\cdot\chi_0+\dots$, so two factors $|\psi_\perp|^2$ are found to appear to the left and right of each operator $v$. In what follows, we include those factors in $v$ and finally write the susceptibility as
\begin{align}
\chi(\rb,\rb')=|\psi_\perp(R,z)\psi_\perp(R',z')|^2\sum_m \chi_m(\omega) \frac{\ee^{\ii m(\varphi-\varphi')}}{2\pi},
\nonumber
\end{align}
where, in virtue of the convolution theorem, we have \[\chi_m=\frac{\chi^0_m}{1-v_m\chi^0_m}\] in terms of the matrix elements
\begin{align}
\chi^0_m(\omega)=\frac{ e^2}{\pi\hbar}\sum_{m'}(f_{m'-m}-f_{m'})\frac{1}{\omega-(\varepsilon_{m'}-\varepsilon_{m'-m})+\ii\gamma} \nonumber
\end{align}
and
\begin{align}
v_m&=\frac{1}{2\pi}\int d^3\rb\int d^3\rb'|\psi_\perp(R,z)\psi_\perp(R',z')|^2\ee^{\ii m(\varphi-\varphi')}\frac{1}{|\rb-\rb'|}. \nonumber
\end{align}
As mentioned above, integration over transversal coordinates $(R,z)$ in $v_m$ cancels the Coulomb singularity at $\rb=\rb'$. For simplicity, instead of considering a specific shape for the cross section of the ring along with the resulting function $\psi_\perp$, we introduce the smoothing effect of such integration through an effective minimum interaction distance $\delta$, so that the induced charged is considered to be distributed along an infinitely thin ring and the Coulomb interaction is replaced by $1/\sqrt{|\rb-\rb'|^2+\delta^2}$, with $\rb$ and $\rb'$ now indicating positions along the ring circumference of radius $a$. This leads to
\begin{align}
v_m\approx 2\int_0^\pi d\varphi\frac{\cos(m\varphi)}{\sqrt{2a^2(1-\cos\varphi)+\delta^2}}. \nonumber
\end{align}
Finally, combining the above results, the polarizability for circularly polarized light reduces to
\begin{align}
\alpha_\pm(\omega)&=-\int d^3\rb\int d^3\rb' \left(\frac{R \ee^{\mp\ii \varphi}}{\sqrt{2}}\right) \left(\frac{R' \ee^{\pm\ii \varphi'}}{\sqrt{2}}\right) \chi(\rb,\rb') \nonumber\\
&=-\pi\left(\int_0^\infty R^2dR\int_{-\infty}^\infty dz \;|\psi_\perp(R,z)|^2\right)^2\;\chi_{\pm1} \nonumber\\
&\approx-\pi a^2\chi_{\pm1}=\frac{-\pi a^2\chi^0_{\pm1}}{1-v_{\pm1}\chi^0_{\pm1}}, \nonumber
\end{align}
where in the last line we take advantage of the narrowness of the ring to approximate $R\approx a$ inside the integrals and further use the normalization of $\psi_\perp$. The calculations presented in the main text for nanorings are carried out for $a=12\,$nm and $b=1\,$nm.

\subsection{Disk}

Following a previous study \cite{paper236}, we model the disk as an infinite potential well of cylindrical shape with radius $a$ in the $x$-$y$ plane and height $h\ll a$ along the normal direction $z$. The electron wave functions are then defined in the $R<a$, $|z|<h/2$ region as $\psi_{lm}(\rb)=A_{lm}A_zJ_m(Q_{lm}R)\sin(\pi z/h)\ee^{\ii m \varphi}/\sqrt{2\pi}$, where $A_{lm}=\sqrt{2}/[aJ_{m+1} (\zeta_{lm})]$ and $A_z=\sqrt{2/h}$ are normalization constants, $Q_{lm}=\zeta_{lm}/a$, and $\zeta_{lm}$ is the $l^{\rm th}$ zero of $J_m$. In this expression, we have introduced the lowest-energy state along $z$, assuming that higher-energy states along that direction can be ignored. We then write the electron energies $\hbar\varepsilon_{lm}=\hbar^2Q^2_{lm}/2\me$ referred to that state. Following a similar procedure as for the ring, we can expand the susceptibility as
\begin{widetext}
\begin{align}
\chi(\rb,\rb')= \frac{4}{h^2}\sin^2\bigg(\frac{\pi z}{h}\bigg) \sin^2\bigg(\frac{\pi z'}{h}\bigg) \sum_{mll'} \chi_{m,ll'}(\omega) A_{lm}J_m(Q_{lm}R)A_{l'm}J_m(Q_{l'm}R')   \frac{\ee^{\ii m(\varphi-\varphi')}}{2\pi}\label{chidisk}
\end{align}
in terms of matrix elements $\chi_{m,ll'}$. Different azimuthal components $m$ are decoupled, so they can be calculated independently according to Eq.\ (\ref{chi}) from the matrices $\chi^0_m$ and $v_{m}$, also expanded in the radial direction using Bessel functions in similar expansions as in Eq.\ (\ref{chidisk}), with $\chi_{m,ll'}$ replaced by $\chi^0_{m,ll'}$ and $v_{m,ll'}$, respectively. More precisely, the matrix element of $\chi^0_m$ reduce to
\begin{align}
&\chi^0_{m,ll'}=\frac{A_{lm}A_{l'm}}{\pi}\frac{e^2 }{\hbar} \sum_{m'l''l'''}(f_{m'-m}-f_{m'})\frac{A_{l''m'}^2 A_{l'''m'-m}^2 }{\omega-(\varepsilon_{m'}-\varepsilon_{m'-m})+\ii\gamma} \nonumber\\
&\times \int_0^a R dR J_{m}(Q_{lm}R)J_{m'}(Q_{l''m'}R) J_{m'-m}(Q_{l'''m'-m}R)\int_0^a R' dR' J_{m}(Q_{l'm}R')J_{m'}(Q_{l''m'}R') J_{m'-m}(Q_{l'''m'-m}R'),\nonumber
\end{align}
while those of $v_m$ are obtained by assimilating the two functions $(2/h)\sin^2(z)$ into the integral along the $z$ direction:
\begin{align}
v_{m,ll'}&=\frac{2A_{lm}A_{l'm}}{\pi h^2} \int d^3\rb\int d^3\rb'\sin^2\bigg(\frac{\pi z}{h}\bigg) \sin^2\bigg(\frac{\pi z'}{h}\bigg) J_{m}(Q_{lm}R)J_{m}(Q_{l'm}R')\ee^{\ii m(\varphi-\varphi')}\frac{1}{|\rb-\rb'|}.\nonumber
\end{align}
Using the identities
\begin{align}
\frac{1}{|\rb-\rb'|}=\frac{1}{2\pi} \int_0^{2\pi} \de \varphi_Q\int_0^{\infty} \de Q  \ee^{\ii Q R \cos(\varphi-\varphi_Q)} \ee^{-\ii Q R' \cos(\varphi'-\varphi_Q)} \ee^{-Q|z-z'|} \nonumber
\end{align}
and
\[2\pi\ii^mJ_m(kr)=\int_0^{2\pi} d\theta\,\ee^{\ii m \theta}\ee^{\ii kr\cos\theta},\]
we can further simplify the matrix elements of $v_m$ as
\begin{align}
v_{m,ll'}&=2\pi A_{lm}A_{l'm}h^2 \int_0^\infty dQ H(Qh) \frac{Q_{lm}a J_{m}(Q a)J_{m}(Q_{lm}a)}{Q^2-Q_{lm}^2} \frac{Q_{lm}a J_{m}(Qa)J_{m}(Q_{l'm}a)}{Q^2-Q_{l'm}^2},\nonumber
\end{align}
where $H(t)=[20\pi^2 t^3+3t^5-32\pi^4(1-e^{-t}-t)]/[4t^2+8\pi^2t]^2$. With these matrix elements, we readily find
\begin{align}
\alpha_\pm(\omega)&=-\int d^3\rb\int d^3\rb' \left(\frac{R \ee^{\mp\ii \varphi}}{\sqrt{2}}\right) \left(\frac{R' \ee^{\pm\ii \varphi'}}{\sqrt{2}}\right) \chi(\rb,\rb') \nonumber\\
&= - \pi \sum_{l,l'} \chi_{\pm1,ll'}(\omega) A_{1l}A_{1l'}\int_0^a R^2 \de R \,J_{1}(Q_{l1}R) \int_0^a R^2 \de R\,J_{1}(Q_{l'1}R)    \nonumber\\
&=- 2\pi a^4  \sum_{l,l'} \frac{\chi_{\pm1,ll'}(\omega)}{\zeta_{1l}\zeta_{1l'}}              \nonumber
\end{align}
\end{widetext}
for the polarizability of a disk under circularly-polarized illumination. The results in the main text for the disk are calculated for $a=12\,$nm and $h=1\,$nm.

\section{Drude model for rotating nanoparticles}

\subsection{Inclusion of noninertial forces}

In the frame rotating with the particle, the total force acting on its electrons is
\begin{align}
{\bf F}=-e\Eb-2\me\vec\Omega\times\vb-\me\vec\Omega\times(\vec\Omega\times\rb),
\nonumber
\end{align}
where the three terms in the right-hand side are the electric, Coriolis, and centrifugal forces, respectively. The latter is conservative and can be described though a potential $U(\rb)=-\Omega^2r_\perp^2/2+U_0$, where $r_\perp$ is the radial distance to the rotation axis. Under such a potential, the equilibrium electron density inside the particle $n(\rb)$ tends to pile up near its outer boundary. We intend to derive an expression for the optical conductivity by using the classical Boltzmann transport equation, which describes the evolution of the electron distribution $f(\rb,\kb,t)$ as a function of position, wave vector, and time. More precisely, we have
\begin{align}
\frac{\partial f}{\partial t}=-\gamma(f-f_0)-\frac{1}{\hbar}\left(e\Eb+2\me\vec\Omega\times\vb\right)\cdot \nabla_{\kb}f, \label{BE}
\end{align}
where the first term introduces a phenomenological inelastic relaxation to the unperturbed distribution $f_0(\rb,\kb)$ at a rate $\gamma$, whereas the right-most term accounts for the force acting on the electron. We note that the latter does not contain the centrifugal force, which is assumed to be described by $f_0$: this distribution is obtained in the presence of the potential $U(\rb)$ and is taken to satisfy the condition $\sum_\kb f_0(\rb,\kb)=n(\rb)$. In Eq.\ (\ref{BE}), we have adopted the local approximation by ignoring a term arising from the spatial gradient of $f$. Nevertheless, the conductivity depends on $\rb$ though the unperturbed electron density $n(\rb)$. Now, we work in frequency space and express the field as $\Eb(t)=\Eb\ee^{-\ii\omega t}+{\rm c.c.}$ This allows us to make the substitution $\partial/\partial t\rightarrow-\ii\omega$ and readily obtain $f(\rb,\kb,\omega)$ from Eq.\ (\ref{BE}). The current density can then be expressed as a sum over electron wave vectors, $\jb^{\rm ind}(\rb,\omega)=(-e\hbar/\me)\sum_\kb\,\kb\,f(\rb,\kb,\omega)$, where we have assumed a free-electron dispersion relation, for which the electron velocity is simply given by $\vb=\hbar\kb/\me$. Putting these elements together and focusing on field and current components along the plane perpendicular to the rotation direction, we find $\jb_\perp^{\rm ind}(\omega)=\sigma(\omega)\cdot\Eb_\perp$, where
\begin{align}
{\bf \sigma}(\omega)=&\frac{\sigma_{\rm D}(\omega)}{1+\beta^2}\;\left[\begin{matrix}
1 & -\beta    \\
\beta   & 1   \\
\end{matrix} \right]
\label{sigmafinal}
\end{align}
is the $2\times2$ local conductivity tensor, in which the rotation velocity appears through the parameter $\beta=-2\ii\Omega/(\omega+\ii\gamma)$, and the factor $\sigma_{\rm D}=\ii \omega_{\rm D}/(\omega+\ii\gamma)$ represents the Drude conductivity in the absence of rotation with $\omega_{\rm D}=(e^2/\me)\,n(r)$ proportional to the local electron density $n(r)$. Repeating the same analysis for graphene (with $\vb=\vF\hat\kb$ having a uniform magnitude given by the Fermi velocity $\vF\approx10^6$m/s), we find exactly the same expressions, but now $\omega_{\rm D}=e^2\EF/\pi\hbar^2$ is determined by the local Fermi energy $\EF$ and the conductivity is 2D rather than 3D. Incidentally, the inclusion of a Lorentz magnetic force along the rotation axis in the above analysis produces an additional term that can be fully absorbed in $\beta=\ii(\omega_{\rm c}-2\Omega)/(\omega+\ii\gamma)$, where $\omega_{\rm c}=eB/m^* c$ is the cyclotron frequency corresponding to the magnetic field $B$, and we have $m^*=\me$ for the free-electron system and $m^*=\EF/\vF^2$ for graphene.

\subsection{Effect of the Coriolis force}
\begin{figure*}[t]
\begin{centering}
\includegraphics[width=0.8\textwidth]{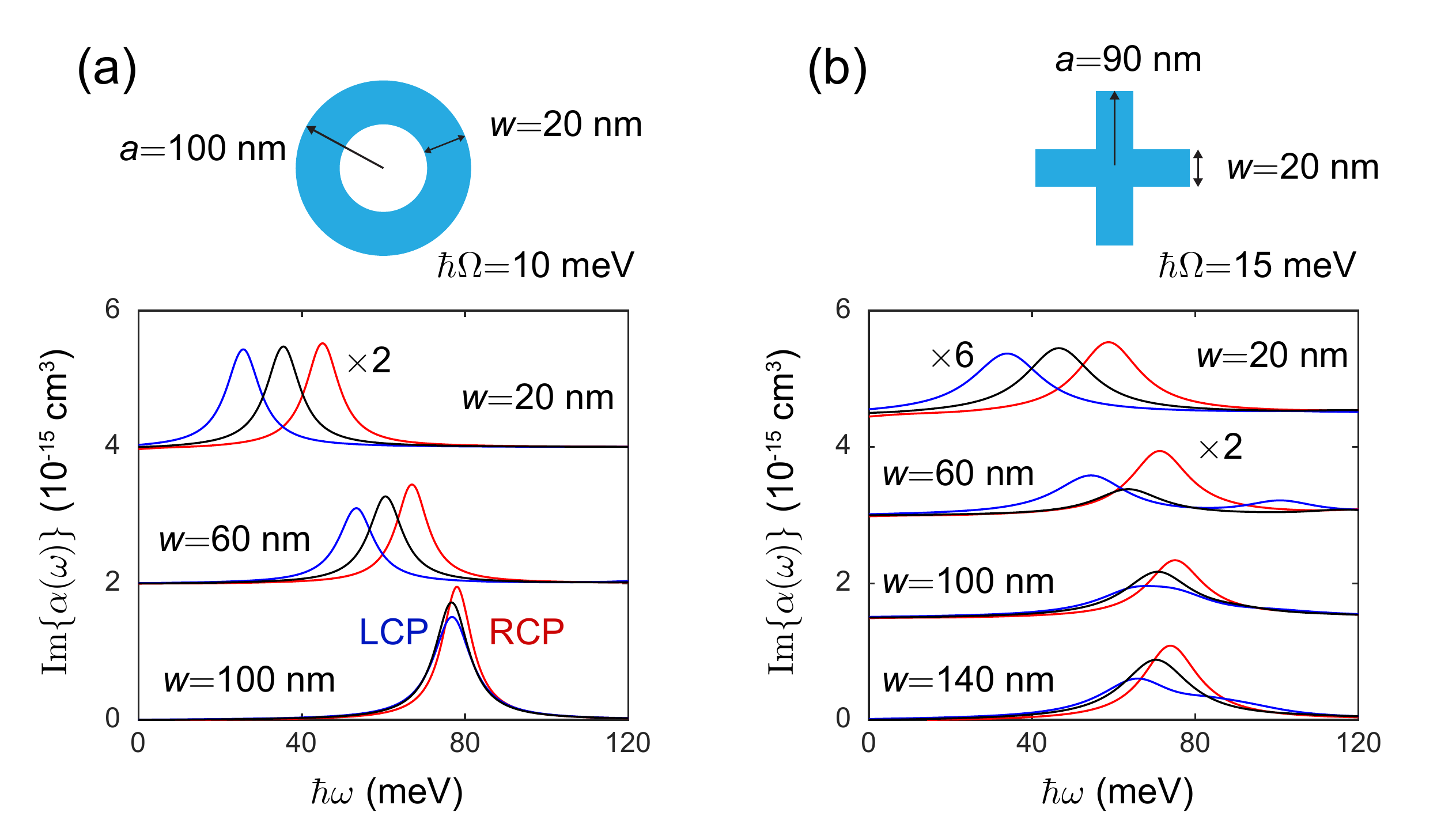}
\par\end{centering}
\caption{Optical response of graphene nanorings (a) and nanocrosses (b) of different widths $w$ rotating at frequency $\Omega$, as calculated within the Drude model by including the effect of the Coriolis force, but ignoring the centrifugal force for the sake of discussion. Red and blue curves correspond to RCP and LCP light, while black curves stand for the motionless particles. We set the rotation frequency according to $\hbar\Omega=10\,$meV and $15\,$meV for disks and crosses, respectively. The graphene is doped to a Fermi energy $E_{{\rm F}0}=0.2$\,eV and we assume an optical damping of $10\,$meV and $20\,$meV for rings and crosses, respectively.}
\label{FigS1}
\end{figure*}
In order to clearly reveal the effect of the Coriolis force on the polarizabilites of rotating particles with different geometries, we first disregard the centrifugal force by assuming a particle electron density $n(\rb)$ independent of rotation velocity. We consider graphene nanoparticles of ring- or cross-like shape, as shown in the upper insets of Fig.\ \ref{FigS1}. We obtain the polarizabilties of these particles by using a finite-difference method to solve Maxwell's equations in the frequency domain, with the local response of the material described through the 2D conductivity tensor (\ref{sigmafinal}). Figure\ \ref{FigS1}(a) shows the polarizabilites of graphene nanorings of different widths rotating at the same frequency $\Omega$ given by $\hbar\Omega=10\,$meV, and calculated in the Drude model. The narrowest ring (width $w=20$\,nm, radius $a=100$\,nm) shows strong CD characterized by a resonance frequency splitting $|\omega_+-\omega_-|\approx 2\Omega$, in agreement with the quantum mechanical results discussed in the main text. This effect becomes weaker with decreasing $w$, and finally the CD almost disappears for a graphene nanodisk ($w=a$) due to the cancellation between shifts induced by the rotational Doppler effect and the Coriolis force. A similar behavior can also be observed for rotating graphene nanocrosses, as shown in Figure \ \ref{FigS1}(b). The CD of the crosses increases when they become narrower, which is consistent with the interpretation presented in Fig.\ 1 of the main text.

\subsection{Effect of the centrifugal force}

We study the effect of the centrifugal force by simulating graphene nanoparticles rotating around an axis perpendicular to the plane of the material and in which the electron density is redistributed due to rotation as prescribed by the Drude model (see below). In general, this requires a numerical solution, but for nanodisks and thin nanocrosses the charge distribution can be found analytically. To this end, we assume zero temperature and use the well-known relation between the Fermi energy and the doping electron density in graphene \cite{CGP09} $\EF(\Rb)=\hbar\vF\sqrt{\pi n(\Rb)}$. Both of these quantities can depend on position $\Rb$ within the plane of the material. We assume a homogeneous charge density $n_0=(E_{{\rm F}0}/\hbar\vF)^2/\pi$ in the motionless graphene particle, determined by the doping level $E_{{\rm F}0}$. In the rotating particle, the Fermi energy depends on position in order to compensate for the potential $U(\rb)$ associated with the centrifugal force (see above). We have
\[\EF(\Rb)=E_0+\frac{1}{2}m^*\Omega^2 R^2,\]
which implies that the electron density $n(R)$ is only a function of the radial distance $R$ for particles of arbitrary geometry. The $\Omega$-dependent constant energy $E_0$ is determined by imposing the conservation of the number of electrons $\int d^2\Rb\, n(R)$. Applying this condition to a disk of radius $a$ (total number of electrons $\pi a^2n_0$), the density becomes
\begin{align}
n_{\rm disk}(R)=n_0\frac{1+B^2 a^2 }{(1-B^2 R^2)^2}.  \nonumber
\end{align}
where $B=\Omega /\sqrt{2}\vF$. Similarly, for a thin nanocross formed by four arms of length $a$ and width $w$, assuming $a\gg w$ (narrow arms), we can approximate $\int d^2\Rb\, n(R)\approx 4awn_0$. Then, the charge density redistribution in the rotating cross reduces to
\begin{align}
n_{\rm cross}(R)=\frac{n_0}{(1-B^2 R^2)}\frac{1}{\left[1-(1-B^2 R^2)(Ba)^{-1}{\rm tanh}^{-1}(Ba)\right]}.  \nonumber
\end{align}
As expected, the charge density in both disks and crosses is higher near the external edges when they are rotating.

\begin{figure*} [h]
\begin{centering}
\includegraphics[width=0.8\textwidth]{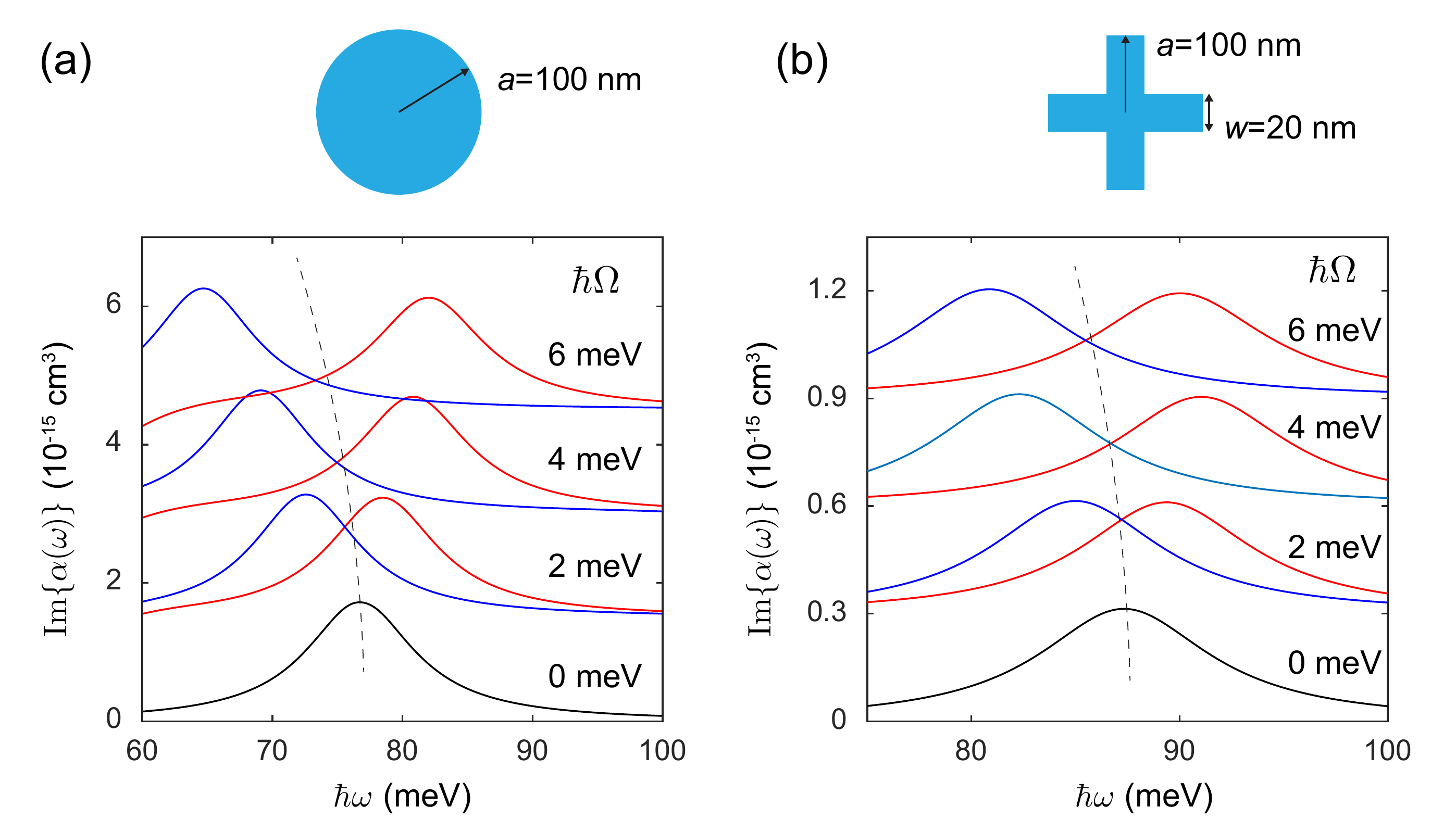}
\par\end{centering}
\caption{Optical response of a rotating graphene nanodisk and nanocross, as calculated within the Drude model including the effect of both Coriolis and centrifugal forces. Red and blue curves correspond to RCP and LCP light, while black curves stand for the motionless particles. Geometrical and rotation frequency parameters are indicated by labels in the plots. We assume a Fermi energy $E_{{\rm F}0}=0.2$\,eV and an inelastic damping of $10$\,meV.}
\label{FigS2}
\end{figure*}

The lowest curve in Fig.\ \ref{FigS1}(a) showed that the Coriolis force cannot induce strong CD in a rotating graphene nanodisk ($a=w=100$\,nm, $\hbar\Omega=10\,$meV) when the centrifugal force was neglected. However, when including the centrifugal force through the redistribution of the electron charge just indicated, strong CD is observed at relatively low $\Omega$, as shown in the calculations presented in Fig.\ \ref{FigS2} for the same graphene disk; interestingly, we further find that the CD of the disk can be larger than that of the ring ($|\omega_+-\omega_-|> 2\Omega$) over a certain frequency range; these results are consistent with the quantum mechanical calculations discussed in the main text. The effect of the quadratic $\beta$ term (see Fig.\ 1 in the main text) can be observed in both disks and crosses, as indicated by the dashed curves in Fig.\ \ref{FigS2}.

\acknowledgments

This work has been supported in part by the Spanish MINECO (MAT2017-88492-R and SEV2015-0522), the ERC (Advanced Grant 789104-eNANO),  the European Commission (Graphene Flagship 696656), AGAUR (2017 SGR 1651), the Catalan CERCA Program, and Fundaci\'o Privada Cellex.

\end{document}